# Multiscale dynamical characterization of cortical brain states: from synchrony to asynchrony

MV Sanchez-Vives[1,2*], A Manasanch[1], A Pigorini[3], A Arena[4], A Camassa[1], BE Juel[4], L Dalla Porta[1], C Capone[5], C De Luca[5], G De Bonis[5], J Goldman[6], M. Sacha[6], A Galluzzi[7], A Pazienti[7], E Mikulan[3], JF Storm[4] PS Paolucci[7], M Massimini[3], M Mattia[7], A Destexhe[6]

[1] Institute of Biomedical Research August Pi i Sunyer (IDIBAPS), Barcelona, Spain; [2] ICREA, Barcelona, Spain; [3] Department of Biomedical and Clinical Sciences, University of Milan, Milan, Italy; [4] Brain Signalling Lab, Division of Physiology, Faculty of Medicine, Institute of Basic Medical Sciences, University of Oslo, Oslo, Norway; [5] INFN Istituto Nazionale di Fisica Nucleare, Sezione di Roma, Italy; [6] Paris-Saclay University, CNRS, Gif sur Yvette, France; [7] Natl. Center for Radiation Protection and Computational Physics, Istituto Superiore di Sanità (ISS), Rome, Italy

Correspondence:

Maria V. Sanchez-Vives, MD, PhD
Institute of Biomedical Research August Pi i Sunyer
Rosello 149-153
08015 Barcelona
msanche3@recerca.clinic.cat

# Abstract

The cerebral cortex spontaneously displays different patterns of activity that evolve over time according to the brain state. Sleep, wakefulness, resting states, and attention are examples of a wide spectrum of physiological states that can be sustained by the same structural network. Furthermore, additional states are generated by drugs (e.g., different levels of anesthesia) or by pathological conditions (e.g., brain lesions, disorders of consciousness). While the significance of understanding brain states in relation to brain dynamics and behavior has become increasingly evident over the past two decades, a unified definition of brain states remains elusive. In this review, we focus on two extremes of this spectrum: synchronous versus asynchronous states. These functional states predominantly underlie unconsciousness and consciousness, respectively, although exceptions exist. Our aim is to integrate data from different levels into a multiscale understanding, ranging from local circuits to whole-brain dynamics, including properties such as cortical complexity, functional connectivity, synchronization, wave propagation, and excitatory-inhibitory balance that vary across states and characterize them. Experimental and clinical data, as well as computational models (at micro-, meso-, and macrocortical levels) associated with the discussed brain states, are made available to readers.

# Introduction: From past to present—the growing interest in spontaneous activity and cortical dynamics

Until the last decade of the twentieth century, the physiological investigation of the cerebral cortex was largely based on the stimulus-evoked response paradigm. Using this paradigm, classical studies of somatosensory (Mountcastle et al., 1957) or visual processing (Hubel and Wiesel, 1962) provided an important basis for modern cortical neuroscience. However, while much of this cortical work focused on stimulus encoding or mapping of receptive fields, few objections were raised against the fact that these experiments were often done under deep anesthesia. Furthermore, stimulus-evoked cortical responses were investigated with a variety of anesthetics and different depths of anesthesia were often used in different laboratories, illustrating a rather weak focus on differences between brain states and their types of spontaneous baseline activity. Decades after the pioneering discovery of the brainstem's ascending reticular activation system by (Moruzzi and Magoun, 1949), a renewed interest in spontaneous activity as a basis for understanding cortical function emerged in the early 1990s. Earlier pioneering studies of neuronal firing during spontaneous activity (Burns and Webb, 1976) were largely descriptive, lacking a paradigm for deeper dynamic and mechanistic insight. The field of sleep physiology was however always focused on spontaneous activity, investigating the structure of spontaneous sleep, and its different phases and wake transitions (Kleitman, 1987; Steriade and Hobson, 1976). Not surprisingly, one of the neuroscientists who strongly contributed to a change in the focus of cortical physiology, identifying the importance of spontaneous activity, was a sleep neurophysiologist: Mircea Steriade (Steriade, 2003; Steriade and Hobson, 1976). In 1993, Steriade, McCormick and Sejnowski (Steriade et al., 1993a) provided an integrative view of the different thalamocortical emergent rhythmic patterns in different brain states, their cellular and network mechanisms, and their potential role in information processing. Since the early 1990s, the value of spontaneous activity as a way of understanding networks and their complexity has increased exponentially. The further development of brain imaging (fMRI), the introduction of novel methods for functional connectivity analysis (Friston et al., 1993), the identification of the default mode network (Raichle et al., 2001) and the growth of resting state studies (Deco et al., 2011), or the current surge in consciousness research (Crick and Koch, 2003; Laureys, 2005; Mashour et al., 2020; Seth and Bayne, 2022; Storm et al., 2024;

Tononi and Edelman, 1998), have further contributed to igniting the interest in spontaneous brain activity.

The term "brain states" refers to different functional modes of the brain, supported by the same structural network (Greene et al., 2023; Kringelbach and Deco, 2020; McCormick et al., 2020; Sporns, 2022). They express different spatiotemporal emergent patterns of activity associated with different modes of information processing (Poulet and Crochet, 2019) and behaviors (Stringer et al., 2019). Brain states can be classified according to arousal levels and activity modes, such as wakefulness and stages of sleep (REM and NREM sleep) (McCormick et al., 2020; Greene et al., 2023; Marin-Llobet et al., 2025), or according to the consciousness level (e.g., awareness and coma) (Laureys, 2005). They can also correspond to various cognitive processes such as focusing on a task, experiencing different emotions, or meditating (Tang et al., 2012). Brain states can also vary under the effects of drugs (e.g., anesthesia) (Arena et al., 2017; Colombo et al., 2019; Farnes et al., 2020; Dasilva et al., 2021; Nilsen et al., 2024), psychedelics (Arena et al., 2022; Bayne and Carter, 2018), or in clinical conditions such as disorders of consciousness (Casali et al., 2013; Laureys, 2005) or stroke (Favaretto et al., 2022). Brain states are typically identified through electrophysiological techniques (e.g., EEG, local field potential (LFP), neuroimaging (e.g., fMRI, calcium or voltage sensitive dyes) which can capture the unique spatiotemporal patterns of neural activity associated with different states. The cerebral cortex spontaneously displays a wide range of spontaneous patterns of activity that correlates with a specific brain state (Buzsáki, 2006). These patterns can range from synchronous to asynchronous (Harris and Thiele, 2011) and involve the integration of mechanisms at various levels of organization, from the microscale, such as ion channels (McCormick et al., 2020), to the dynamics of the entire brain network (Lynn and Bassett, 2019). However, understanding these different brain states from a multiscale perspective remains a significant challenge in neuroscience.

In this review article, we aim to provide a multiscale exploration of different brain states, ranging from synchronous to asynchronous. We include brain states expressed by the cortical network, drawing from animal models, human data, and computational models. While reviews often integrate already published figures, here we also include "live figures" online, which contain panels that provide access to data, simulations, or even allow active generation of figure panels, all through the EBRAINS platform. In doing so, we offer readers the tools to become active experimenters by making our own data and models available within the online figures.

# A synchronous state disconnected from the world: slow oscillations

## Slow oscillations as the default mode of the cortical network

Our lives oscillate between periods of sleep and wakefulness, but where is the baseline activity of our cerebral cortex? We cannot truly speak of a steady state or baseline activity, as we are continuously influenced by circadian rhythms (Ruan et al., 2021). However, we can refer to a default activity pattern of the cortical network—a common pattern toward which cortical circuits tend to converge under different conditions, as we will discuss in the following sections. This default activity pattern is the cortical slow oscillation (Sanchez-Vives and Mattia, 2014).

Pathological cortical activities, such as epileptic discharges or cortical spreading depression, represent the most synchronized cortical events. However, among physiological activities, the slow oscillatory rhythm is the most synchronized activity of the cerebral cortex. Being synchronized means involving large populations of neurons, resulting in large-amplitude signals. For this reason, slow oscillations are more easily detectable and were reported in the first EEG recordings in humans by Hans Berger (Berger, 1929), due to their large amplitude and detectability. A detailed description of this rhythmic activity at the cellular level came 64 years later (Steriade et al., 1993b). In that study, slow oscillations were characterized as rhythms under 1 Hz, consisting of active periods interspersed with silent periods (Figure 1A). Active periods were found to be supported by synaptic activity that, when suprathreshold, induced neuronal firing. These periods were followed by silent phases marked by absence of firing. The terms active and silent periods have largely been replaced by Up and Down states in the literature, although the terminology can vary across laboratories.

Steriade et al. described slow oscillations as being of cortical origin, since they could be recorded even after surgical disconnection of the cortex from the thalamus, in isolated gyri, and even in disconnected cortical slabs (Timofeev, 2000). The sufficiency of cortical circuits to generate fully developed slow oscillations has also been supported by several other studies, some of which we include in this review. Slow wave sleep (SWS) is defined by slow oscillations and delta waves which occur at a frequency of 0.5–4 Hz in humans (Amzica and Steriade, 1998; Dang-Vu et al., 2008). Figure 1B displays slow waves from a rat during SWS, while being recorded from different cortical areas (Torao-Angosto et al., 2021). When the same subject, with the same electrodes' locations, is placed under deep anesthesia (Figure 1B), the cerebral cortex also expresses slow oscillations. The similarity between the slow waves during SWS and during anesthesia depends on the sleep stage, on the anesthetic being used and on the anesthesia level. There are many similar properties: for example, the waves share similar cortical origins and preferentially propagate along the mesial components of the default network, although anesthesia-induced waves have not been reported to effectively entrain spindle activity (Alkire et al., 2008; Murphy et al., 2011).

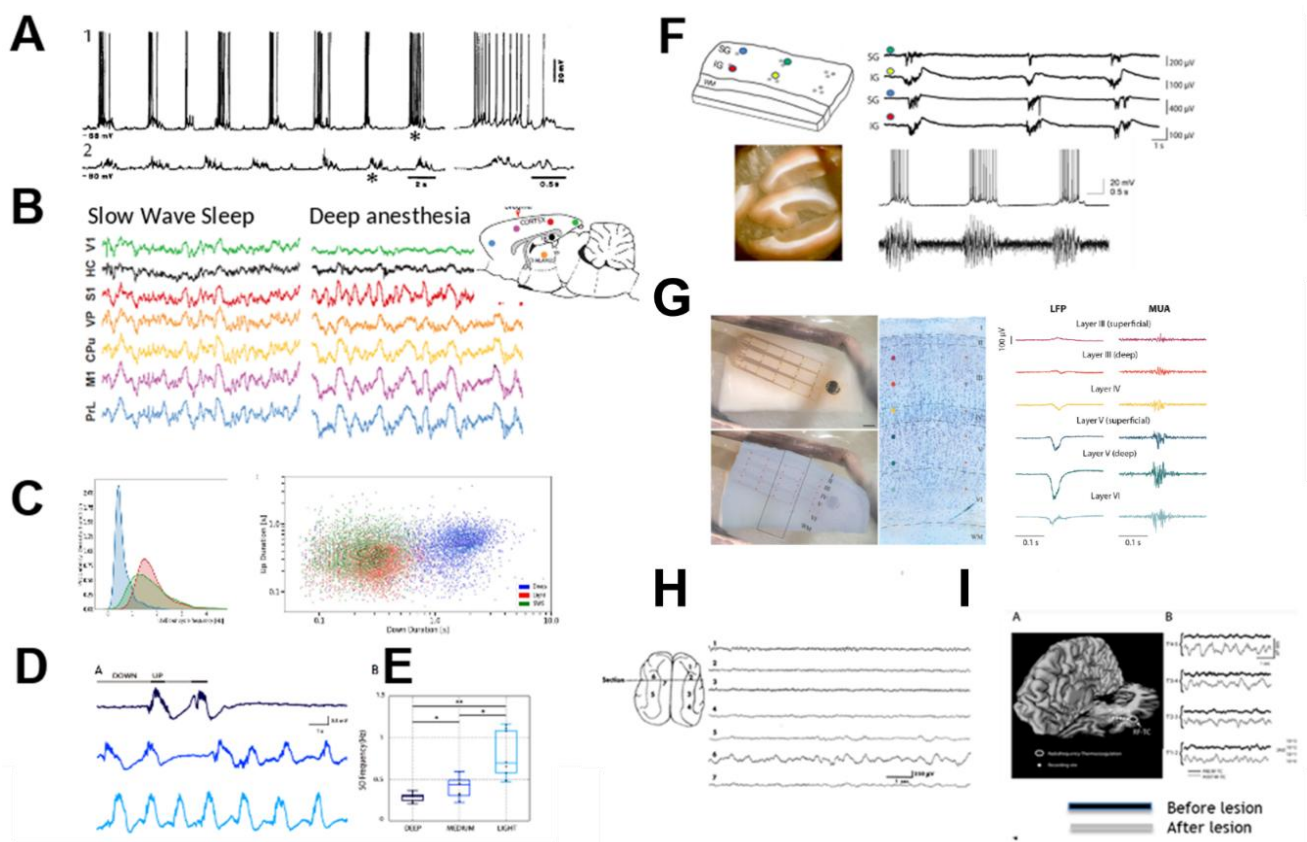

***Figure 1 Multiscale exploration of neocortical slow (<1 Hz) oscillations across preparations, species and conditions.** **(A)** Intracellular recordings (suprathreshold, top; subthreshold, bottom) from cat neocortex under ketamine–xylazine anesthesia, illustrating spontaneous alternations between depolarized "Up" states and hyperpolarized "Down" states. Scale bars: 2 s (horizontal), 10 mV (intracellular). Adapted from (Steriade et al., 1993b). **(B)** Example local field potential (LFP) traces recorded during natural slow-wave sleep (left) and*

*deep anesthesia (right) from multiple sites: secondary somatosensory cortex (S1), primary visual cortex (V1), ventral posterolateral thalamus (VP), primary motor cortex (M1) and prelimbic cortex (PrL). Despite differing brain states, Up/Down alternations are observed throughout cortical and thalamic structures. Adapted from (Torao-Angosto et al., 2021).* **(C)** *Left: Probability density functions of Up state (solid lines) and Down state (dashed lines) durations under deep (blue), medium (green) and light (red) isoflurane anesthesia, showing a progressive lengthening of Up states with lighter anesthesia. Right: Scatter plot of individual Up vs. Down durations, revealing distinct clusters corresponding to each anesthesia level. Adapted from (Torao-Angosto et al., 2021)* **(D)** *Representative LFP traces recorded simultaneously from superficial (top), middle (middle) and deep (bottom) anesthesia levels in anesthetized rat. The bar above the traces indicates Down (left) and Up (right) states. Scale bars: 0.5 s, 500 µV. Adapted from (Dasilva et al., 2021).* **(E)** *Box plots of slow-oscillation frequency (Hz) measured under deep, medium and light anesthesia (n = 6 rats per condition). Oscillation frequency increases significantly with lighter anesthesia (p<0.01, one-way ANOVA with post-hoc Tukey). Adapted from (Dasilva et al., 2021).* **(F)** *In vitro laminar recordings in acute ferret cortical slices. Left: Schematic of a linear electrode array spanning supragranular (SG), granular (G) and infragranular (IG) layers. Left below: Photograph of a cortical slice in the recording chamber. Center right: Nissl-stained section showing electrode track and contact positions (colored circles). Right: Simultaneous intracellular (upper) and extracellular (lower) traces from SG (black/green) and IG (gray/red), demonstrating propagating Up–Down transitions in vitro. Scale bars: 1 s; 20 mV (intracellular), 100 µV (extracellular). Adapted from (Capone et al., 2019b) and (Sanchez-Vives and McCormick, 2000).* **(G)** *Laminar multisite silicon-probe recordings from rat somatosensory cortex in vivo. Left: Photographs of the cortical surface showing probe position before (top) and after (bottom) DiI coating. Middle: Nissl-stained coronal section with DiI track and contact locations (red dots). Right: Layer-specific LFP (left column) and multi-unit activity (MUA; right column) traces from layers III (superficial), II (deep), IV, V (superficial/deep) and VI. Note the slight lead of superficial layers in initiating Up transitions. Scale bars: 0.1 s; 100 µV (LFP), 0.1 s. Adapted from (Dasilva et al., 2021)..* **(H)** *Extracellular field recordings from cat sensorimotor cortex before and after focal ischemic stroke. Left: Schematic of recording sites (1–7) overlaid on cortical sectors. Right: LFP traces showing that slow oscillations persist in peri-infarct cortex post-stroke. Scale bar: 200 µV, 1 s. Adapted from (Gloor et al., 1977).* **(I)** *High-density EEG in human patients before (blue) and after (gray) a focal cortical lesion (Sarasso et al., 2020). Left: three-dimensional reconstruction of the lesioned hemisphere with electrode positions (white circles).*

The quantification of properties of Up and Down states in SWS versus deep and light anesthesia revealed that SWS was indeed closer to the properties of light anesthesia, while deep anesthesia displayed significantly longer Down states (Torao-Angosto et al., 2021; Figure 1C). SWS is a spontaneous and physiological brain state, as other sleep or awake stages are. Because they are spontaneous, as are the transitions between them, these emergent patterns are hard to control experimentally. In order to systematically investigate the different dynamic patterns that can be expressed by the cortical network, one experimental approach is to gradually vary anesthesia levels, which allows the system to visit different states. Some of these may not be spontaneously expressed physiologically; however, their exploration provides valuable dynamical information about the underlying network, providing elements for computational modelling (Bettinardi et al., 2015; Cimenser et al., 2011). By using this strategy, we can vary anesthesia levels and investigate the spectrum of slow oscillations (Figure 1D; (Dasilva, 2021)). Different anesthesia levels result in a modulation of the frequency of slow oscillations or Up and Down states (0.25–4 Hz; Tort-Colet et al., 2021; Manasanch et al., 2025), which reflect different states of the network also reflected in functional complexity, wave propagation, or network complexity (Bettinardi et al., 2015; Arena et al., 2022; Dasilva, 2021). Therefore, slow oscillations are one dynamical regime that contains heterogeneity within, some of which we revise in this article.

Slow oscillations emerge in the cortex in situations such as SWS or anesthesia, therefore in the cortex *in situ*. Interestingly, almost-identical patterns—both in intracellular and multiunit features—have been observed in cerebral cortex slices *in vitro* (Figure 1F; (Sanchez-Vives and McCormick, 2000)). Slow waves *in vitro* start in cortical layer 5 (Sanchez-Vives and McCormick, 2000), as has been reported *in vivo* (Beltramo et al., 2013; Chauvette et al., 2010; Sakata and Harris, 2009), and also propagate as waves across the cortical tissue as they do *in situ*. This finding is highly revealing that the necessary building blocks for slow waves to emerge are already contained in the local cortical circuits. That isolated cortical circuits tend to generate slow waves is also evidenced by their emergence in the perilesional tissue that has thus suffered a disconnection (Gloor et al., 1977; Russo et al., 2021; Sarasso et al., 2020) (Figure 1G,I), contributing to the idea of slow waves as the default activity pattern of the cortical network (Sanchez-Vives et al., 2017).

## Role of the thalamus in slow oscillations

Overwhelming evidence supports the cortical origin of slow oscillations and the fact that cortical circuits are sufficient to generate slow oscillations (Sanchez-Vives and McCormick, 2000; Timofeev, 2000). However, the cortex is reciprocally connected with other subcortical nuclei, notably the thalamus (Jones, 2012; Steriade et al., 1993a; Timofeev and Chauvette, 2013). Because the thalamus is a highly rhythmic structure, and a well characterized generator of patterns such as spindle waves (Contreras and Steriade, 1996; Krosigk von et al., 1993; Lüthi and Nedergaard, 2025; Sanchez-Vives and McCormick, 1997), its recurrent connectivity with the cortex has an influence on cortical slow oscillations. The findings that some thalamic nuclei, such as the thalamic reticularis nuclei, could generate under certain *in vitro* experimental conditions (e.g., activation of metabotropic glutamate receptors) a slow (<1Hz) rhythm, supported the idea that slow oscillations were of thalamic origin (Blethyn et al., 2006), an idea that reverberated in the field for some time. More recent studies in human epilepsy suggested that the anterior thalamus precede neocortical slow waves during NREM sleep in these patients (Schreiner et al., 2022), and computational models have argued that coordinated "cardinal oscillators" in thalamus and cortex may both be needed to sustain *in vivo* slow oscillations (Crunelli and Hughes, 2009).

Nonetheless, we hold that slow oscillations are fundamentally cortical: they persist virtually unchanged when subcortical inputs are severed (Timofeev et al., 2000) and can emerge locally during wakefulness when cortical areas become transiently deafferented as a consequence of brain lesions (Gloor et al., 1977; Sarasso et al., 2020). In intact brains and physiological conditions, however, thalamocortical interactions can modulate the timing or spatial propagation of cortical activity depending on the brain states as a result of the reciprocal connectivity.

## How slow are slow oscillations?

Although the slow oscillations were first described as a <1Hz rhythm, its frequency range is still a matter of debate that we will not elude in this review. As argued above, the cortical network, or even local cortical circuits, have all the necessary mechanisms to generate slow oscillations. This has been fully demonstrated in extreme cases in which the cortical circuits are fully isolated, such as cortical slices or cortical slabs (Sanchez-Vives and McCormick, 2000; Timofeev et al., 2000; Covelo et al., 2025); and even cultures of cortical cells express some "network events" (Eytan and Marom, 2006). The minimum mechanisms for slow oscillations to occur are basically recurrent connectivity and adaptation mechanisms (see computational evidence below), which would result in a certain frequency range (Compte et al., 2003; Mattia and Sanchez-Vives, 2012). For example, variations in excitability by changing extracellular potassium around physiological levels (~4mM), result in a variation in slow wave frequency and regularity (Sancristobal et al., 2016). *In situ*, the cerebral cortex is recurrently connected across cortical areas as well as with subcortical nuclei, and, importantly, with the thalamus. Even when these subcortical nuclei may not initiate the slow oscillations, their own intrinsic properties and rhythmicity may modulate the resulting cortical rhythm. Furthermore, this modulation varies depending on the brain states. These different factors result in a broadening of the frequency range of slow oscillations, which is indeed wider than the original proposal (<1Hz) of Steriade et al (Steriade et al., 1993b).

Varying levels of anesthesia, a paradigm to vary brain states in a controlled way (e.g., (Cimenser et al., 2011)) revealed that at light levels of anesthesia, we can still record slow oscillations, albeit at higher frequencies that generally assumed, right over 1 Hz (Figure 1C; (Dasilva, 2021)) all the way up to 2–4 Hz (Tort-Colet et al, 2021). These final frequencies enter the delta frequency range. So, should slow oscillations include delta frequencies? Slow oscillations and delta have been classically classified as two distinct neurophysiological phenomena with different spatial and temporal properties (Steriade et al., 1993b). In the current work we propose that slow oscillations and delta rhythms are indeed a continuum, with the same underlying mechanisms, but under different states and/modulatory influences. As mentioned, it is not only the average frequency that varies, but also the regularity of the oscillation (Mattia and Sanchez-Vives, 2012; Sancristobal et al., 2016; Tort-Colet et al., 2021). We propose that the coefficient of variation of slow waves may be suggestive of the mechanisms at play. Sancristobal et al. (2016) manipulated cortical network excitability by progressively changing the extracellular potassium concentration. Interestingly, slow oscillations *in vitro* showed the least variability (minimum coefficient of variation and collective stochastic coherence) at 4 mM K, with variability increasing at both lower and higher potassium -and excitability- levels. While highly regular slow waves *in vivo* suggest a single mechanism of generation, namely local cortical circuits, irregular cycles are compatible with a larger role of subcortical nuclei, albeit they can be computationally justified by higher (or lower) levels of cortical excitability. Interestingly, slow oscillations and delta waves have been reported to have competing roles in sleep-dependent memory consolidation (Kim et al., 2019). The involvement of different cortical areas and subcortical nuclei can form the basis of these different roles in memory.

## Defining neuronal bistability: silent periods (Down states or Off-periods) as determinant of cortical dynamics

During Up states, periods of sustained depolarization and recurrent firing within the slow wave regime, local cortical circuits maintain persistent activity through tightly balanced excitation and inhibition (Compte et al., 2009; Haider et al., 2006). The intervals of quiescence that follow, terminologically have been called Down states, silent periods, and in multiunit/EEG studies, Off-periods. All denote the hyperpolarized, quiescent part of the slow oscillation, detectable across recording modalities and species, their duration directly sets the frequency of slow oscillations, with longer Down states yielding slower rhythms (Camassa et al., 2022; Timofeev et al., 2000).

Multiple, overlapping mechanisms contribute to terminating Up states and enforcing the subsequent Down states. These encompass incoming excitatory inputs (Haider et al., 2006; Shu et al., 2003); it has been proposed that short-term synaptic depression depletes excitatory drive as network activity persists, precipitating a transition to silence synaptic depression ((Bazhenov et al., 2002) but see (Benita et al., 2012)). Withdrawal of tonic thalamic excitation—or "disfacilitation"—could further hyperpolarize cortical neurons at Up state termination (Contreras and Steriade, 1996). The slow afterhyperpolarization recorded intracellularly during Down states (e.g., figure 7C in (Contreras et al., 1996; Sanchez-Vives and McCormick, 2000)) indicates that prolonged $K^+$ conductances likely play a role both in terminating Up states and in sustaining Down states. Several specific $K^+$-dependent mechanisms have been proposed, while extracellular $K^+$ buildup during periods of firing further shifts ionic gradients to activate the $Na^+/K^+$-ATPase to favor afterhyperpolarization, contributing to the the silent period (Fröhlich et al., 2006; Sancristobal et al., 2016). Activity-dependent $K^+$ currents, including $Na^+$-dependent and $Ca^{2+}$-dependent hyperpolarization, gradually accumulate during prolonged firing due to intracellular rising of sodium and potassium, driving neurons into Down states (Sanchez-Vives and McCormick, 2000; Wang et al., 2003). The duration of these Down states is positively correlated with the firing frequency of the preceding Up states (Sanchez-Vives et al., 2010). Metabotropic $GABA_B$ receptor activation adds an inhibitory brake that prolongs Down states and sharpens the Up-Down transition (Barbero-Castillo et al., 2021a; Mann et al., 2009; Perez-Zabalza et al., 2020), while ATP-sensitive $K^+$ channels link network excitability to metabolic state (Cunningham et al., 2006). Interestingly, $K^+$-mediated afterhyperpolarization are suppressed by neuromodulators such as acetylcholine and noradrenaline—agents that facilitate the sleep-to-wake transition (Brumberg et al., 2000; Dalla Porta et al., 2023; Madison and Nicoll, 1986; Pedarzani and Storm, 1993; Schwindt et al., 1988)—thereby providing a means to disrupt the bistable regime upon arousal.

Far from being functionally inert, Down states serve key roles in cortical dynamics. By separating successive Up states, they reset synaptic and neuronal conditions (Reig and Sanchez-Vives, 2007) effectively gating information flow and interrupting ongoing patterns of activity across local and long-range networks (Camassa et al., 2022). Detailed analyses of *in vivo, in vitro*, and *in silico* data reveal that these silent epochs exhibit a metastable architecture—beginning with a highly synchronized, deterministic phase and transitioning into a more stochastic, desynchronized period—whose balance determines the timing and responsiveness of the next Up state (Camassa et al., 2022).

When Up states are evoked—whether by sensory, electrical, or TMS perturbation—the duration of the subsequent silent period provides a sensitive index of the network's bistability and complexity. In healthy and pathological human studies, TMS-EEG measurements show that Off-periods disrupt causal interactions and correlate inversely with the perturbational complexity index (PCI; (Casali et al., 2013; Rosanova et al., 2018)), offering a window into the state-dependent capacity of cortical circuits to sustain integrated activity (Massimini et al., 2024). Similar results have also been obtained in rats, suggesting that large-scale effective connectivity is suppressed after Off-periods, in a slow oscillatory regime (Arena et al., 2021).

## The wave nature of slow oscillations

Slow oscillations collectively emerge as travelling waves of activity across the cortex. This phenomenon has been described for human cortical activity during slow wave sleep (Massimini et al., 2004; Nir et al., 2011), during anesthesia in both humans (Lewis et al., 2012; Murphy et al., 2011) and animal models (Chauvette et al., 2011; Dasilva, 2021; Liang et al., 2021; Mohajerani et al., 2010; Pazienti et al., 2022; Ruiz-Mejias et al., 2011; Sheroziya and Timofeev, 2014; Stroh et al., 2013), and even across cortical slices *in vitro* (Capone et al., 2019b; Covelo et al., 2025; Sanchez-Vives and McCormick, 2000; Wester and Contreras, 2012). The speed of propagation of these waves varies in all these cases, but it is in the range of 4–50 mm/s. Indeed, cortical networks can be viewed as a field of local cell assemblies coupled via both short-range lateral connections and long-range callosal cortico-cortical connectivity. From a physics standpoint, the cortical network can be seen as an excitable medium composed of spatially organized interacting oscillators (Ermentrout and Kleinfeld, 2001). In this context, travelling waves emerge as a change of activity in one local assembly eliciting in turn a response in the nearby networks giving rise to a kind of chain reaction propagating in space. In these types of excitable systems, the waves can also be "stationary," leading to synchronized activity where all the oscillators in the network evolve with the same phase. This activation mode has also been observed in the intact brain (Benucci et al., 2007; Bolt et al., 2022).

## Diversity of waves

Travelling slow waves are a hallmark of spontaneous activity of the cerebral cortex, not just *in vivo*, but also when kept isolated as in brain slices maintained *in vitro* (Capone et al., 2019b; Covelo et al., 2025; Sanchez-Vives and McCormick, 2000; Wester and Contreras, 2012) as shown in Figure 2B. In the intact brain, slow waves preferentially initiate in the frontal cortex and propagate backward to posterior (rostral) cortical areas (Massimini et al., 2004; Mohajerani et al., 2010; Nir et al., 2011; Ruiz-Mejias et al., 2011; Sheroziya and Timofeev, 2014; Stroh et al., 2013) (see Figure 2A). However, there is not a limited repertoire of propagation modes. Indeed, slow waves appear to be influenced by the previous sensory experience (Han et al., 2008) and their occurrence is sometimes a kind of stochastic rehearsal of previously encoded sensory responses (Luczak et al., 2009). Furthermore, changes in the global brain state can widen such repertoire (see next section) giving rise to non-trivial propagation modes like rotating (i.e., spiral) waves (Huang et al., 2010; Muller et al., 2018). Remarkably, the diversity of slow waves seems to be related not only to differences in the spatiotemporal patterns of propagation. This is rather apparent in the wide variability of measured speeds of propagation across species and recording modalities. On the one hand, analyses of scalp EEG in humans (Massimini et al., 2004) and LFP from ECoG both in human and non-human primates (Davis et al., 2020; Halgren et al., 2019) filtered in the alpha frequency band (7–13 Hz), consistently report velocities of slow waves ranging from 0.5 to 5 m/s, similarly to that found in monkeys with voltage sensitive dyes (Muller et al., 2018) and from neuronal spiking activity in sleeping humans (Nir et al., 2011). On the other hand, smaller velocities have been found in anesthetized rodents. They range from 10 to 40 mm/s both if the activation wavefronts are measured from MUA and single-cell membrane potentials (Fucke et al., 2011; Pazienti et al., 2022; Ruiz-Mejias et al., 2011; Sheroziya and Timofeev, 2014) and if they are estimated from calcium and voltage sensitive dyes imaging (Greenberg et al., 2018; Stroh et al., 2013). Whether these differences are related to having observed different species or because different phenomena have been inspected is still an open question. Analysis tools capable of providing a common workbench to make a fair comparison are now more urgent than ever (Gutzen et al., 2024).

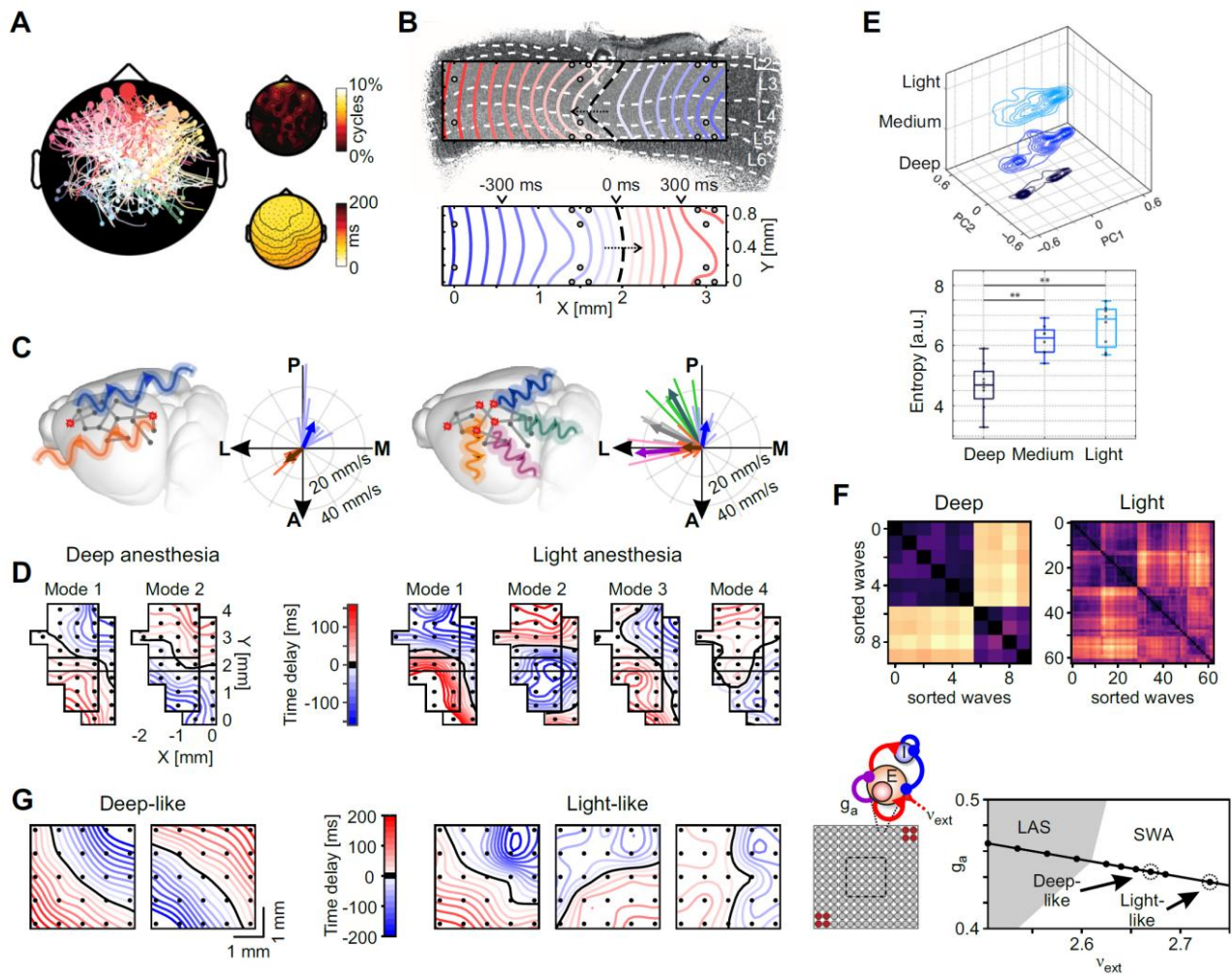

*Figure 2. Cortical travelling waves in vivo, in vitro and in silico.* **(A)** *Slow oscillations in humans from EEG recording occurs as propagating oscillatory patterns whose centers of mass follow trajectories moving from the front to the back of the brain (dot size indicating the origin of the travelling wave). Right-top, map of wave origin, highlighting higher densities in anterior scalp regions. Right-bottom, average wavefront at different delays further showing the rostro-caudal direction of slow-wave activity across the scalp. (Adapted from Massimini et al., 2004).* **(B)** *Average spatiotemporal propagating patterns of activation waves (traveling wavefronts, thick lines coloured by time delay) in a cortical slice clustered in two different "modes of propagation". Top, back ground example cortical slice with electrode position (circles) (Adapted from Capone et al., 2019b).* **(C)** *Schematic representation of the wave propagation modes in the mouse brain for different anesthesia levels. Right, statistics of modes velocity for an example animal are colored according to the results of k-means clustering of traveling wavefronts (i.e., rows of the time lags/delays matrix). Thick arrows, average directions and velocities of each mode. Left and right panels, "Deep" and "Light" anesthesia, respectively.* **(D)**

*Average traveling wavefronts at different time delay for the modes singled out in (C), highlight the increasing complexity of slow-wave activity as anesthesia fades out (C and D, adapted from Pazienti et al., 2022).* ***(E)*** *(Top) Distributions of traveling wavefronteach represented as a point in the ($PC_1$,$PC_2$) plane under three different levels of anesthesia placed along the z-axis. (Bottom) Shannon entropy of wavefront distributions increasing from "Deep" to "Light" anesthesia levels ( Wilcoxon signed-rank test with Benjamini–Hochberg corrections: \* p<0.05, \*\* p<0.01) (Adapted from Dasilva et al., 2021).* ***(F)*** *Similarity matrix of traveling wavefront ordered according to the clustering provided by the dendrograms leaves for Deep (left) and Light (right) anesthesia levels in (C) of an example animal. High and low wavefront similarities, dark and light colors, respectively (Adapted from Camassa et al., 2021).* ***(G)*** *Left panels, average travelling wavefronts in spiking neural network simulation clustered in different "modes of propagation" for Deep-like (left) and Light-like (right) anesthesia level of an example simulation. Center, schematic representation of the simulated network, consisting of a lattice of local excitatory-inhibitory assemblies of leaky integrate-and-fire neurons with adaptation (LIFCA). Anesthesia fading is modelled by increasing the network excitability (stronger excitatory input $v_{ext}$ and reduced adaptation strength $g_a$). Right, bifurcation diagram showing parameter change as a linear trajectory moving from a phase where the network prevalently stays in a low-firing asynchronous state (LAS, burst-suppression-like) and the slow-wave activity (SWA) state. Arrows, parameter combinations used in the left panels (Adapted from Pazienti et al., 2022).*

## Waves as biomarkers of the cortical state

Far from being a monolithic dynamical mode, slow waves display a relatively wide spectrum of spatiotemporal patterns depending on the anesthesia levels (Arena et al., 2021; Dasilva et al., 2021; Liang et al., 2023; Pazienti et al., 2022; Shimaoka et al., 2017) and likely on the different sleep stages. Changes in the diversity of wave propagation modes as anesthesia fades out in the arousal process can inform about the evolution of the functional organization of the network leading to wakefulness and hence underlying the onset of consciousness.

Wave propagation investigated in recent experiments by recording the electrophysiological cortical activity from surface micro-electrocorticographical (µECoG) multi-electrode arrays (MEA) is probed with sufficiently high spatiotemporal resolution. From these recordings, the rise in complexity of slow-wave activity results in apparently once-similar propagating waves being grouped together (Figure 2C-D). Under deep anesthesia the clustered propagation modes are mainly from front-to-back and vice-versa (Figure 2D-left panels) (Greenberg et al., 2018; Pazienti et al., 2022). Complexity increases under light level of anesthesia when direction of propagation is mainly from front to back (Figure 2D-right panels), and their distribution is wider (Liang et al., 2023; Pazienti et al., 2022). Besides, an increase of the speed of propagation has been found with the approach to wakefulness (Figure 2C-right, see the magnitude of the arrows). The increased repertoire of slow waves is even more apparent from the similarity matrices between single waves at different levels of anesthesia (Figure 2F), displaying an increasing degree of randomness directly associated with an increase of entropy of the distribution of propagation modes (Figure 2E, F). Notably, all these changes are suggestive of a rise in the excitability of the cortical tissue as the awake state is approached.

A question then arises: which is the mechanistic origin of such excitability modulation? Simulations of model networks of spiking neurons capturing the detailed statistics of slow waves has been developed to address this open issue (Figure 2G). The models capable of quantitatively replicating experimental observations appear to be in a specific sweet spot of the parameter space (Figure 2G-right). The 'latitude' (Y) and 'longitude' (X) of such a 'geographic' map is determined by two key features of local cell assemblies associated with the amount of (synaptic/metabotropic) excitation they receive and the adaptation level of their spiking frequency. As a result, an intriguing prediction from such models is that changes in the local excitability of cortical assemblies are sufficient to explain how cortical slow waves evolve as anesthesia fades out without requiring any modification of the network connectivity.

# Computational models of slow oscillations

As described above, experimental evidence shows that slow oscillations in the cortex exhibit a wide range of propagation patterns and speeds, modulated by brain state or by anesthetic depth. These diverse dynamics suggest complex, state-dependent changes in cortical excitability. To uncover the mechanisms behind such variability, computational models are essential. They allow controlled exploration of how local parameters—such as synaptic drive and spike-frequency adaptation—can reproduce and predict the observed diversity without altering network connectivity. By bridging experimental data and theory, models help distill the minimal physiological ingredients underlying cortical dynamics.

| Reference Author, year, [Figure.panel /section in the present paper] | Mean Field or Spiking simulation, (Neuron Type) | Topology/ Extension | Brain states/Dynamics | Specific features |
|---|---|---|---|---|
| Latham et al., 2000<br>Tabak et al., 2000<br>van Vreeswijk and Hansel, 2001<br>Holcman and Tsodyks, 2006<br>Gigante et al., 2007<br>Parga and Abbott, 2007<br>Destexhe, 2009<br>Mattia and Sanchez-Vives, 2012<br>Jercog et al., 2017<br>Zerlaut et al., 2018<br>Levenstein et al., 2019<br>Perez-Zabalza et al., 2020 | Mean Field and spiking | Single Population | Fixed points and limit cycles | Spike Frequency adaptation, short-term synaptic depression and slow GABA-ergic transmission |
| Bazhenov et al., 2002<br>Compte et al., 2003<br>Hill and Tononi, 2005<br>Krishnan et al., 2016 | Spiking (HH) | Cortical and thalamo-cortical networks | SW, transition to asynchronous | Metabotropic, Ionotropic and synaptic modulation |
| Mattia and Sanchez-Vives, 2012<br>Tort-Colet et al., 2021<br>[Figure 3A] | Mean Field and Spiking (LIF) | Single population | Deep / Light Anesthesia, SO | Transition from deep to light anesthesia. |
| Goldman et al., 2021<br>Sacha et al., 2024<br>Sacha et al., 2025 | Mean field & Spiking, (AdEx) | Connectome-bases network of cortical areas | SO / Asynchronous | Human, Monkey, Mouse. PCI measures |
| Capone et al., 2019b<br>Barbero-Castillo et al. 2021a<br>[Figure 3B] | Spiking (LIFCA, HH) | Cortical slices | SW | Connectivity as waveguides. Synch. vs asynch. PCI measures. |
| Di Santo et al., 2018<br>Pazienti et al., 2022<br>Pazienti et al., 2024<br>[Figure 2G] | Mean Field and Spiking (LIFCA) | Cortical fields (2D) of local assemblies. | SW, Asynchronous | Synchronous to asynchronous transition and brain criticality. |
| Rulkov et al 2004<br>Ali et al 2013<br>Pastorelli et al., 2019 | Spiking (LIFCA) | 2D grid RS-FS, exponential connectivity decay | SW, asynchronous | Large scale, 70G synapses |
| Markram et al., 2015<br>Figure 3C | anatomically realistic neurons | 2D anatomically realistic model (slice) | Synchronous (SO) / Asynchronous | SO in biologically accurate neurons and slice |
| Capone et al., 2023<br>Figure 3D | Mean Field (AdEx) | Whole mouse hemisphere, 2D model | Anesthesia, SW, asynchronous, spiral waves | Inferred from experimentally recorded activity |
| Esser et al., 2007<br>Wei et al., 2018<br>Capone et al., 2019a<br>Golosio et al., 2021<br>Luca et al., 2023 | Spiking (AdEx), Plastic | Thalamo cortical, RS-FS, TC-RE single or multi-area | SO, NREM, Awake | Cognitive and energetic functions of sleep |

**Table 1. Computational models of slow oscillations.**

In the investigation of slow waves, computational models serve as an essential tool to complement experimental approaches. Despite the apparent simplicity of slow wave activity, limitations in experimental interventions—such as the ability to test only a finite number of drug concentrations, intensity amplitudes, or other experimental manipulations—restrict our ability to fully dissect the mechanism at the micro-circuitry level. Furthermore, slow wave activity generates complex phenomena at meso- and macroscales. Much of the complexity arises from the recurrent connectivity of the network, which renders the experimental dissection of participating mechanisms nearly inaccessible. For instance, activity-dependent adaptation mechanisms play a crucial role. However, due to the circularity of the system, cause and effect become recurrently interdependent. Consequently, the resulting dynamics are highly nonlinear, creating a system that can readily exhibit paradoxical and counterintuitive behaviors. This underscores the significance of employing computational models. Models may capture different levels of neurophysiological measures (from intracellular to whole-brain dynamics), they can cover different spatial scales, or be organized according to different topologies, or express different brain states. The topology of the connectivity can model the interaction in single populations of neurons, 2D slices or 3D regions. Table 1 proposes a categorization of a few representative models that are shown in Figure 3.

Local cell assemblies in the cerebral cortex can be described as two homogeneous and interacting populations of excitatory (glutamatergic, regular-spiking) and inhibitory (GABAergic, fast-spiking) neurons (Beltramo et al., 2013; Burns and Webb, 1976; Compte et al., 2009; Douglas and Martin, 2004; Eytan and Marom, 2006). They spontaneously express a wide repertoire of dynamical regimes giving rise to metastable states and coherent oscillations observed both *in vivo* and *in vitro* cortical networks (Andrillon et al., 2021; Arena et al., 2021; Benita et al., 2012). Such richness is faithfully reproduced by model networks of integrate-and-fire (IF) neurons, and mean-field theories offer a compass to predict their collective behavior (Cakan et al., 2022; Destexhe, 2009; di Volo et al., 2019; Gigante et al., 2007; Holcman & Tsodyks, 2006; Mattia & Sanchez-Vives, 2012).

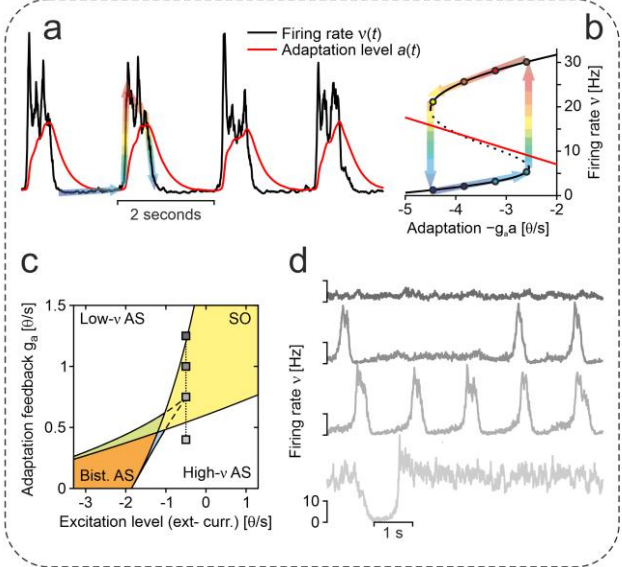

**A** SO mechanism in models

**B** SO in COBA models

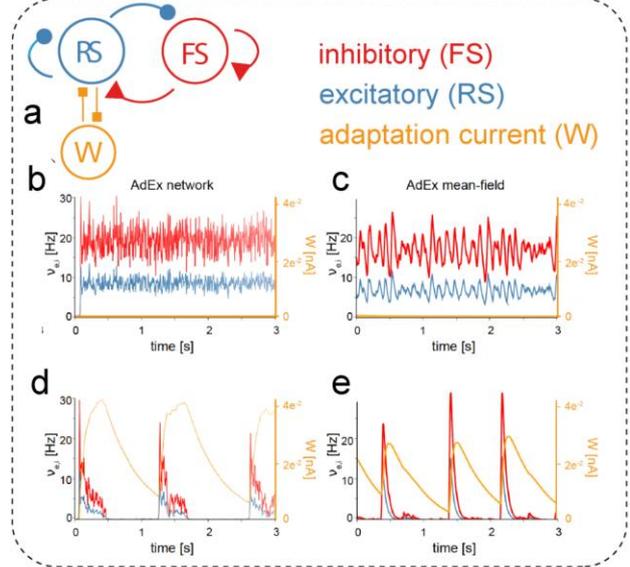

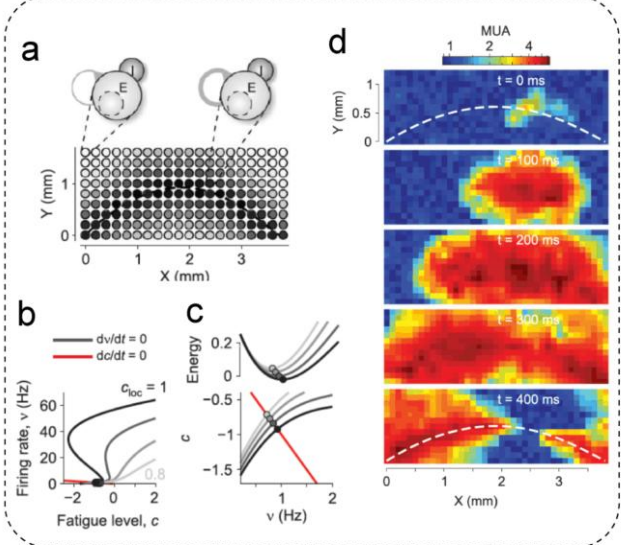

**C** SO in planar models

**D** SO in planar HH models

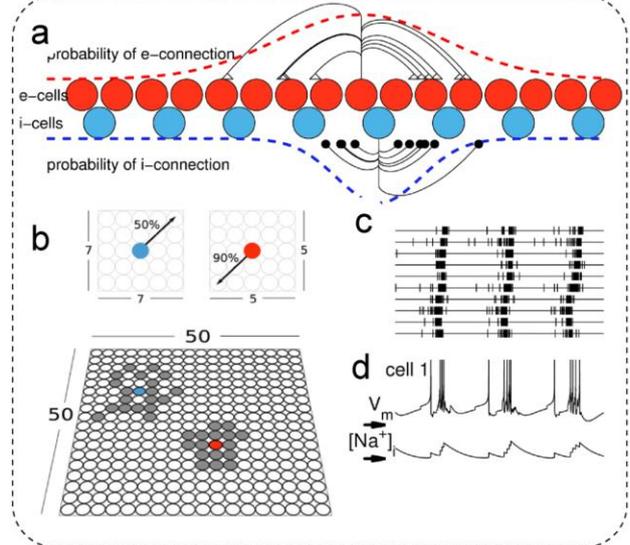

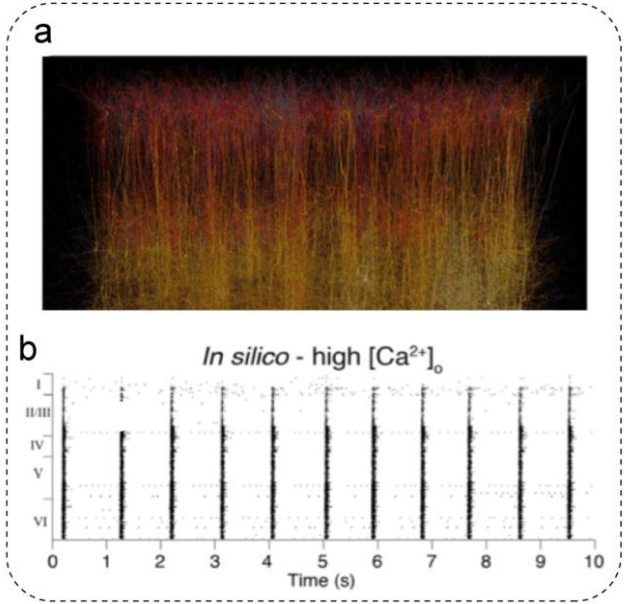

**E** SO in highly detailed models

**F** live tool for dynamics exploration

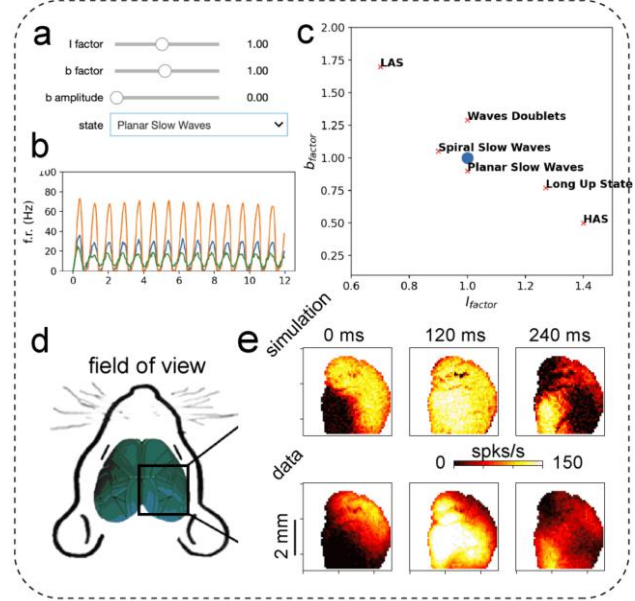

***Figure 3. Computational models of Slow Oscillations and Waves. (A)*** *(Adapted from (Mattia and Sanchez-Vives 2012)) SOs as relaxation oscillators.* ***a.*** *Experimental MUA (black) (equivalent to firing rate $\nu(t)$ in model) and reconstructed adaptation variable a(t).* ***b.*** *Phase plane (a, v) with nullclines and fixed points. Arrows and circles trace the orbit of the relaxation oscillator.* ***c.*** *Excitation-adaptation bifurcation diagram (external excitatory input $I_{ext}$ vs. adaptation strength $g_a$) showing four regimes: LAS, HAS (low/high-firing asynchronous states), and SO (slow oscillations, i.e., limit cycle) and bistable AS (coexistence of LAS and HAS). Vertical dashed line: trajectory modeling the awakening from sleep/anesthesia with sample activity traces in (**A-d**).* ***d.*** *Example firing rates from simulations modeling the awaking across burst-suppression, slow-oscillations and the desynchronized state (from dark to light traces, parameters as in **A-c**).* ***(B)*** *(Adapted from Di Volo et al. 2019) Asynchronous and synchronous dynamics in AdEx networks. a. Network architecture of excitatory RS (blue) and inhibitory FS (red) AdEx neurons. b–c. Mean firing rates in asynchronous and synchronous regimes, differing only in injected SFA. d–e. Time courses of population firing rates and adaptation current $W_e$.* ***(C)*** *(Adapted from (Capone et al., 2019b) Mean-field (MF) networks with spatial structure. a. 2D multimodular arrangement with varying local/global connectivity; darkest circles indicate highest connectivity. b. Nullclines and fixed points in the (v, c) plane across connectivity levels. c. Energy landscapes around fixed points (top), and zoomed phase-plane (bottom), both color-coded as in a. d. Time-lapse snapshots of Up wavefront propagation across the cortical slice. White dashed line: most excitable region (MCS); color: simulated MUA.* ***(D)*** *(Adapted from (Compte et al., 2003) and (Barbero-Castillo et al., 2021a) Hodgkin-Huxley (HH) neuron model in 2D cortical network. a. Schematic of spatial connectivity for pyramidal (blue) and inhibitory (red) neurons, with example connectivity distributions. b. 50×50 neuron grid with 50% local excitatory (7×7) and 90% local inhibitory (5×5) connections. c. Simulated spontaneous MUA activity. d. Intracellular voltage ($V_m$) and sodium concentration in three pyramidal cells.* ***(E)*** *(Adapted from (Markram et al., 2015) Digital reconstruction of neocortical microcircuit. a. Virtual slice with seven unitary microcircuits showing spontaneous activity. b. Rastergrams from random neurons under in vitro- and in vivo-like conditions.* ***(F)*** *Interactive simulation interface. A. UI for selecting parameters: external current, adaptation, oscillatory modulation amplitude. B. Time courses of firing rates from three sample populations and full-network activity (color-coded). C. Parameter space plot of external current and adaptation, including preset regimes. D. Sequential frames of slow-wave propagation in data and simulations.*

Slow oscillation dynamics are a general feature of cortical computational models, since they can be observed consistently in very different systems and simulations. The main mechanism consists of a combination of two ingredients: the recurrent connectivity, which maintains the excitability in the network, and adaptation, an activity-dependent fatigue variable that provides inhibitory feedback (Barbero-Castillo et al., 2021a; Bazhenov et al., 2002; Compte et al., 2003; Destexhe, 2009; di Volo et al., 2019; Jercog et al., 2017; Holcman & Tsodyks 2006; Latham et al., 2000; Levenstein et al., 2019; Mattia and Sanchez-Vives, 2012; Parga & Abbott, 2007; Tabak et al., 2000; van Vreeswijk & Hansel, 2001). The interplay between these two forces results in different stable states, such as Up and Down states (Figure 3A). We will describe this computational canonical circuit in some details since it supports slow oscillations and brain state transitions.

The metastable activity spontaneously expressed by these circuits is not capable of accounting for some of the characteristic features of slow oscillations. Indeed, the hopping between Up and Down states due to the activity fluctuations is not random. Rather, Up-Down oscillations are quasi-periodic and their rhythm is too slow to be captured by the characteristic timescales of the mentioned of excitatory-inhibitory circuits (Beltramo et al., 2013; Compte et al., 2003; Jercog et al., 2017; Lim & Rinzel, 2010; Mattia & Sanchez-Vives, 2012; Levenstein et al., 2019). The noise-driven escape from metastable Up and Down states needs an additional key ingredient: spike-frequency adaptation. Indeed, sustained firing in nervous cells is known to adapt at lower levels displaying a kind of fatigue mechanism. This phenomenon can be due to both changes in ionic concentration-dependent potassium ($K^+$) currents (Benda and Herz, 2003; Compte et al., 2003; Gigante et al., 2007) (spike frequency adaptation) or depletion of synaptic resources (Holcman & Tsodyks, 2006; Hill & Tononi, 2005; Tabak et al., 2011) (short-term depression). Both changes are activity-dependent and well described by an adaptation variable which rises and falls during the Up and Down periods in related fatigue and recovery phases, respectively. Adaptation displays relatively extended time scales ranging from a few hundreds of milliseconds to several seconds and modulates the stability of Up and Down states leading to the quasi-periodic alternation observed in experiments.

EIA (excitatory-inhibitory with adaptation) cortical circuits then display the typical slow-fast dynamics of relaxation oscillators (Latham et al., 2000; Gigante et al., 2007; Liam & Rinzel, 2010; Mattia & Sanchez-Vives, 2012). In their (adaptation, firing rate) state space, this canonical circuit follows a closed orbit, well described by mean-field theories and fully captured by spiking neuron models (see Table 1). As described in the main text (Figure 3B), key parameters like adaptation and synaptic strength can induce the destabilization of slow oscillations giving rise to awake-like and burst-suppression-like states further highlighting the predictive power of this EIA canonical circuit (Fig. 3A). At the same time, these two forces determine a parameter space that includes not only synchronous patterns (slow oscillations), but also asynchronous ones (as in wakefulness), and even relatively silent spaces (as in pathological states like a coma). This simple mechanism can also be used to account for the dynamics observed in different anesthesia levels (Destexhe 2009; Mattia & Sanchez-Vives 2012; Tort-Colet et al., 2021).

Similar dynamical regimes can be observed in a module of Adaptive Exponential Integrate-and-Fire model (AdEx), with conductance-based neurons both in the spiking simulation and the mean field description (Figure 3B) (Augustin et al., 2017). When these modules are arranged on a planar structure, slow oscillations propagate as travelling waves across the 2D network (Figure 3C) (Ali et al., 2013; Capone et al., 2019b; Pazienti et al., 2022; Rulkov et al., 2004; Zerlaut et al., 2018). Slow waves can also be modelled using recurrently connected Hodgkin and Huxley neurons (Barbero-Castillo et al., 2021a; Bazhenov et al., 2002; Compte et al., 2003) forming in a 2D network (Figure 3D). Adaptation is then modelled by realistic potassium dependent currents, shaping the alternance between Up and Down states. The planar network allows the estimation of spontaneous or evoked wave propagation under different conditions, such as different levels of inhibition (Barbero-Castillo et al., 2021a). In such models, we can further obtain information about the evolution of ionic currents activation, intracellular ionic levels, or modify these and other parameters concerning synaptic connections or connectivity patterns.

Although networks of integrate-and-fire neuron models and related dynamic mean-field theories may appear as a rough approximation of the details and heterogeneity observed in biologically inspired networks, their collective behavior captures the richness of global dynamics of cortical networks. Indeed, even more detailed models designed to include neurons reconstructed from *in vitro* staining procedures and incorporating Hodgkin-Huxley voltage- and concentration-gated ionic conductances (Figure 3E) (Markram et al., 2015) display quantitatively similar features of the slow rhythms spontaneously expressed by mean field models of the cortical network. As expected from simple models and related theories, varying [$Ca^{2+}$]-dependent potassium currents underlying spike-frequency adaptation leads to a modulation of the statistics of slow oscillations, like the mean frequency and the coefficient of variation of the Up-Down cycle.

Here we provide an interface to interactively explore simulations and different regimes of cortical dynamics (Figure 3F). In Figure 3 panel F is linked to a Jupyter Lab environment, where the user can interactively change the simulation parameters and observe in real-time the emergence of different dynamical regimes, from slow-wave propagation patterns to an asynchronous regime. The simulation stands for a columnar mean-field model equipped with lateral connections inferred from experimentally acquired cortical activity. The simulations run almost instantaneously, allowing for rapid visualization of system dynamics in response to parameter tuning, such as external input current, adaptation strength, and the amplitude of the oscillatory neuromodulation. Moreover, users can choose from a list of preset parameters showing specific dynamics: planar slow waves, spiral slow waves, doublets of waves, the alternation between microarousals and slow waves, high-frequency waves and others. The interface is available on the EBRAINS KG, on the EBRAINS Collab and on Github (https://t.ly/hMNTJ).

## Slow Oscillation and the Unconscious state

Slow oscillations reflect a bistable dynamic—alternating between Up and Down states (here referred to as Off-periods)—that occurs not only at the cellular level but also at the macroscopic level, as shown in both human and animal studies. In humans, neuronal Off-periods can be detected noninvasively by analyzing scalp EEG in the time-frequency domain. Simultaneous EEG and intracranial recordings in animal models and in patients undergoing presurgical evaluation have shown that neuronal hyperpolarization during Off-periods corresponds at the scalp to a suppression of high-frequency activity (>20 Hz) and to the negative deflection of slow oscillations or delta waves (<4 Hz) (Mukovski et al., 2007). This negative peak is flanked by Up states or periods of high frequencies—including sleep spindles—creating the graphoelement or a distinct EEG waveform pattern known as the "sleep slow wave," which appears during both non-REM sleep and anesthesia. These time- and frequency-domain features characterize not only stable delta-wave patterns but also isolated events such as spontaneous and evoked K-complexes (Cash et al., 2009). Moreover, the rate of these EEG events correlates directly with the depth of sleep and with the level of anesthesia (Dasilva, 2021; Silber et al., 2007; Warnaby et al., 2017).

Crucially, the presence of such graphoelements—and the underlying cortical bistability—disrupts functional network interactions in structurally intact circuits, as demonstrated by recordings in deep sleep and under anesthesia. Functional MRI and EEG studies reveal altered brain dynamics, and combined TMS-EEG recordings show that, unlike the complex, deterministic cortical response to a TMS pulse during wakefulness, non-REM sleep and anesthesia evoke only a simple, sleep-like slow wave (Casarotto et al., 2016). This profound reduction in brain complexity—driven by slow oscillations and cortical bistability—has been linked not only to the loss of consciousness in sleep and anesthesia (induced by different drugs) but also to disorders of consciousness in brain-injured patients (Casarotto et al., 2016). For example, patients in a vegetative state (also called Unresponsive Wakefulness Syndrome) exhibit slow-wave activity during wakefulness: they open their eyes, maintain basic reflexes, and cycle through sleep and wakefulness, yet show no signs of awareness. Recent work suggests that their impaired brain function may result from an excessive intrusion of sleep-like dynamics into an otherwise awake brain (Rosanova et al., 2018). Interestingly, Up and Down states have also been recorded in the cortex during focal limbic seizures, suggesting that a state of decreased cortical arousal may contribute to mechanisms of loss of consciousness during seizures (Yue et al., 2020; Gummadavelli et al., 2021).

Furthermore, numerous experiments indicate that the boundary between wakefulness and sleep—and more generally between synchronous and asynchronous brain states—is more fluid than once thought (Sarasso et al., 2014a). Sleep-like dynamics can intrude, to varying degrees and in localized regions, into wakefulness (and vice versa), as well as into dream-like psychedelic states (Arena et al., 2022; Li and Mashour, 2019). Local slow waves during wakefulness also appear after sleep deprivation (Vyazovskiy et al., 2011) disturbing task performance in mice, or spontaneously during lapses of attention in humans (Andrillon et al., 2021).

# From synchrony to asynchrony: from unconsciousness to consciousness

Cortical dynamics can be understood as ranging from highly synchronized, bistable regimes—characteristic of deep non-REM sleep and surgical anesthesia—to the desynchronized, high-frequency activity observed during wakefulness. In the synchronized extreme, populations of neurons undergo near global Up state depolarizations followed by Down state hyperpolarizations, producing large-amplitude slow waves in the EEG and suppressing high-frequency oscillations. This rhythmic alternation reduces interareal communication (Nir et al., 2011), favoring local synchrony over the rich, spatially distributed interactions that underlie conscious processing (Horovitz et al., 2009). As a consequence, perturbations such as TMS pulses elicit only a simple, stereotyped slow wave, reflecting a collapse of causal interactions across the cortical network (Massimini et al., 2007).

Transitioning from sleep toward wakefulness—or emerging from anesthetic suppression—destabilizes the bistable attractor that dominates cortical dynamics during slow waves. Increases in neuromodulatory drive (e.g., acetylcholine, noradrenaline) reduce the duration and depth of Down states, allowing parts of the cortex to maintain depolarized, high-frequency firing (Steriade, 2001). At the macroscopic level, this shift manifests as a flattening of the EEG spectral slope and a widening of the power spectrum (Höhn et al., 2024). Functionally, these changes release a repertoire of dynamic states: network interactions become both richer and more differentiated, the brain's effective connectivity expands, and causal perturbations give rise to deterministic cascades rather than isolated slow waves.

The progression from unconsciousness to consciousness can be sudden, but can also be gradual, through progressively longer microarousals, brief periods of awakening during slow wave sleep (Lo et al., 2004; Lüthi and Nedergaard, 2025; Parrino et al., 2012) or deep anesthesia (Manasanch et al., 2025; Tort-Colet et al., 2021). By quantifying the balance of synchronous and asynchronous features—through metrics such as perturbational complexity, spectral slope, or travelling-wave entropy described here—we can track an individual brain's state, which can have clinical applications for anesthesia depth, diagnosing disorders of consciousness, and also for investigating the neural substrates of consciousness.

# Getting out of the slow oscillatory attractor: awakening

The slow oscillations or the cortical default activity pattern (Sanchez-Vives and Mattia, 2014) is at full display in states such as NREM sleep and deep general anesthesia, while they globally dominate cortical activity and there is a complete disconnection from the external world or unconsciousness. Markers of the default state can then be measured at multiple scales: as bistable neuronal firing at the cellular level, as slow waves in LFP at the circuit level, and as shifts in the spectral slope of EEG recordings at the whole brain level. However, as the arousing neuromodulatory input from the brainstem and thalamus increases, cortical circuits and neurons tend to transition out of the slow oscillatory regime (Steriade et al., 1993b). This transition is associated with measurable signatures across scales: single cells fires in an asynchronous manner and slow waves in LFPs become less frequent (Vyazovskiy et al., 2009); a more asynchronous, high-frequency, low-amplitude activity becomes dominant (Barbero-Castillo et al., 2021a; Nilsen et al., 2024), and the spectral slope of the EEG flattens (Colombo et al., 2019; Lendner et al., 2020; Nilsen et al., 2024; Zhang et al., 2023). Ultimately, this transition promotes a more reliable effective connectivity across scales, therefore facilitating the propagation of information within the cortical network (Bettinardi et al., 2015; Arena et al., 2021; Hönigsperger et al., 2024).

At the single cell level, the default activity pattern is associated with bistable neural firing patterns, alternating between synchronized Down states with hyperpolarization of the neuronal membrane potential and Up states with depolarization and firing (Mukovski et al., 2007; Steriade et al., 1993b; Volgushev et al., 2006). During awakening from NREM sleep, the membrane potential gradually stabilizes at more depolarized values and with less synchronous firing of action potentials (Figure 4A) (Mukovski et al., 2007; Steriade et al., 1993b). Similarly, during several forms of general anesthesia, cortical circuits and neurons become bistable due to changes in neuromodulatory input and more direct effects of anesthetic drugs on cortical circuits. These effects are reversed during recovery from anesthesia (for example with sevoflurane, Figure 4B). However, as the animal awakens from their disconnected state, the bistable firing gradually disappears, and is replaced by more continuous, higher-frequency, but asynchronous firing of action potentials (Figure 4A,B).

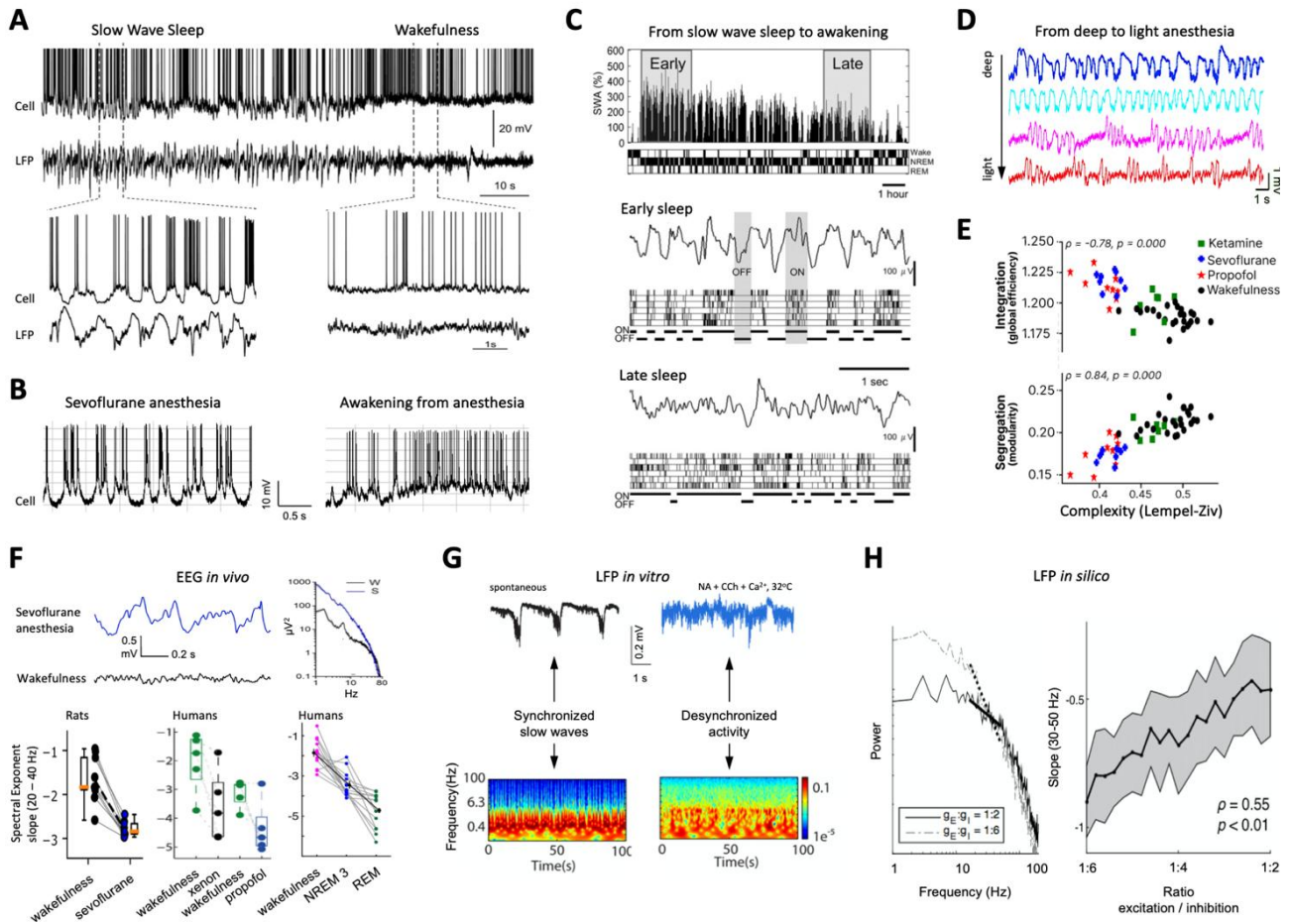

***Figure 4. Emerging from slow waves to wakefulness:*** *(**A**) Micro-scale: single cell membrane potential; Up states and Down states from in vivo intracellular recording of a cat cortical neuron during slow wave sleep, awakening and wakefulness (from (Mukovski et al., 2007)). (**B**) In vivo patch clamp recording from a rat cortical neuron during sevoflurane anesthesia and awakening (from C. Hönigsperger, A. Arena & J.F. Storm (2019, unpublished)). Also shown is simultaneous recording of the local field potential (LFP) from the cat. (**C**) Meso scale: MUA, LFP; On and Off periods. (Top) decreasing incidence of slow waves in (bottom) LFP recordings and unit activity from rats during early and late sleep, and towards period of wakefulness (from (Vyazovskiy et al., 2009). (**D**) LFP recordings during decreasing depth of general anesthesia with ketamine + medetomidine (from (Tort-Colet et al., 2021)). The Off periods (neuronal silence) become less frequent and duration of On periods (firing of action potentials) increases from early to late sleep stages. A similar effect is seen by decreasing the level of general anesthesia. (**E**) Macroscale: ECoG and EEG. At the ECoG/EEG level from rats, the emergence from anesthesia to wakefulness is characterized by a reduced signal integration (less globally synchronized) and more segregation, therefore increasing in complexity (from (Nilsen et al., 2024)). (**F**) (Top) Intracranial EEG recordings from a rat during general anesthesia (sevoflurane 2.6%) and wakefulness, with relative periodogram (top right) illustrate how the activity changed from low to high frequency during the transition*

*from general anesthesia to wakefulness (from (Arena et al., 2021), This redistribution of frequency powers from slow waves to desynchronized and high frequency activity is characterized by a higher slope (or spectral exponent) of the periodogram in range 20–40Hz. This feature is consistent across species (rats vs humans) and across conditions (wakefulness vs general anesthesia with various drugs, and sleep (from (Arena et al., 2021; Colombo et al., 2019; Lendner et al., 2020)). **(G)** A similar redistribution of power from low to high frequency was also observed in cortical slices in vitro, after bath-application of norepinephrine and carbachol (a cholinergic agonist) with physiological $Ca^{2+}$ concentration and temperature to mimic wakefulness. This caused a switch from a bistable dynamic (Up and Down states) to a more depolarized state with asynchronous firing (from (Barbero-Castillo et al., 2021a). **(H)** A cortical network model showed similar changes of the slope of the periodogram (in a similar high frequency range) associated with a shift in the ratio between excitatory/inhibitory synaptic currents (from (Gao et al., 2017).*

At the cortical circuit level, the bistable neuronal dynamic of NREM sleep and anesthesia produces spontaneous slow waves that can be seen from LFP recordings, (Figure 4A,C,D) (Brown et al., 2010; Nir et al., 2011; Torao-Angosto et al., 2021; Vyazovskiy et al., 2009). In contrast, upon awakening, EEG and LFP signals become progressively more dominated by high frequencies and low amplitude activity, reflecting tonic depolarization with more asynchronous spike firing and synaptic activations (Figure 4A,C) (Brown et al., 2010; ; Manasanch et al. 2025; Mukovski et al., 2007; Steriade et al., 1996; Vyazovskiy et al., 2009). That said, it has been shown in rats that the rhythmic alternation between On- and Off periods (Up and Down states) is not constant throughout the NREM sleep, but depends on the sleep pressure (Figure 4C) (Vyazovskiy et al., 2009). Thus, immediately after a long period of continuous wakefulness, when the sleep need is high, the Off-periods were particularly long-lasting and frequent, and the LFP was mainly characterized by high-amplitude slow waves. Conversely, when approaching awakening, both the duration and the frequency of Off-periods were gradually reduced, and the duration of On-periods increased. A similar dynamic was seen in rats when the depth of general anesthesia decreased towards wakefulness (Dasilva, 2021; Tort-Colet et al., 2021; Manasanch et al., 2025).

At the macroscale (Figure 4E,F), awakening from NREM sleep or general anesthesia is associated with a transition from EEG activity dominated by slow, high-power oscillations, to a sustained, high frequency activity, with reduced low-frequency power. This pattern is consistent across species (Figure 4E) (Arena et al., 2021; Brown, 2010; Lendner et al., 2020; Nilsen et al., 2024), and increase the complexity of the EEG signal (Figure 4E), due to reduced similarity across cortical areas (Nilsen et al., 2024). A similar redistribution of power from low to high frequency can even be seen *in vitro*, in cortical slices in conditions intended to mimic wakefulness (Figure 4G) (Barbero-Castillo et al., 2021a). Generally, the observed modulation of spontaneous activity during awakening is reflected into a more positive slope (or spectral exponent), in the high frequency range (20–40 Hz) of the power/frequency plot, when comparing wakefulness to general anesthesia, in both rats and humans, and wakefulness to sleep in humans (Figure 4F) (Arena et al., 2021; Colombo et al., 2019; Lendner et al., 2020; Nilsen et al., 2024). Importantly, a similar change in the slope of the periodogram was found to be associated with the modulation of the ratio between excitatory/inhibitory synaptic currents in the LFP from in an *in silico* model of a cortical network (Gao et al., 2017) (Figure 4H), suggesting that the changes observed as animals awaken can be indicative of a shift in the balance between excitation and inhibition in the underlying networks (Gao et al., 2017). This hypothesis is further supported by mean field models *in silico*, where an increment of excitatory "external current" ($I_{ext}$) contributed to push the dynamical regime of the mean-field model from synchronous slow oscillations toward high frequency and desynchronized activity, as described by the bifurcation diagram (Tort-Colet et al., 2021). In the same mean-field model, the network state was also modulated by a self-inhibiting adaptation process, which contributed to maintaining the neuronal activity in a slow oscillatory mode.

A related hypothesis for the transition between brain states emphasizes the control of neuronal states by modulatory action by the ascending activation system (Moruzzi & Magoun, 1949). This is supported by classic results from in vivo experiments (Steriade et al., 1993a), where bath application of ACh caused neurons *in vitro* to switch from an alternation between Up and Down states to a more depolarized state with regular firing, resembling the *in vivo* changes from slow wave sleep to wakefulness (Dalla Porta et al., 2023). Besides, activation of a photo-switchable muscarinic agonist modulated the cortical slow oscillations both *in vitro* (in mouse cortical slices) and *in vivo* (in mice), by increasing the frequency of the slow-wave activity and increasing the power of high-frequency activity (Barbero-Castillo et al., 2021b). Moreover, injection of the cholinergic agonist carbachol in the prefrontal cortex of rats *in vivo* was sufficient to awaken rats from general anesthesia, with concomitant reduction of slow wave EEG activity (Pal et al., 2018).

Thus, the transition from states associated with hallmark activity of the default cortical state (bistable dynamics and slow waves) can be observed at multiple scales and can be explained by alterations in neuronal state of activation.

# Whole brain modelling of the synchronous to asynchronous transition

Classically referred as the "desynchronized EEG" condition (Niedermeyer and da Silva, 2005; Steriade, 2003), the activity in awake and conscious mammalian brains consists of irregular firing activity and very reduced neural synchrony. Indeed, when the animal attends stimuli, the EEG typically desynchronizes (Crochet and Petersen, 2006; Tan et al., 2014), as well as during active sensing (Harris and Thiele, 2011). Waking up from natural sleep also displays a transition from slow-wave activity to desynchronized activity with irregular firing of cortical neurons (Steriade et al., 2001).

The asynchronous/synchronous transition can also be seen in whole-brain models, as shown in Figure 5 for three species. The simulations used the AdEx mean-field model which was implemented in The Virtual Brain (TVB) (Goldman et al., 2023, 2020). In the AdEx mean-field, each node can potentially display asynchronous-irregular (AI) activity or slow oscillations with Up/Down states, by adjusting the strength of spike-frequency adaptation. When placed in TVB, a network of such nodes can display either asynchronous states or synchronized slow waves. This is seen for the mouse brain (Figure 5A), the macaque monkey brain (Figure 5B) and the human brain (Figure 5C). It is remarkable that when the nodes are set to the AI mode (low adaptation strength), they do not synchronize, while when they are in the Up/Down mode (strong adaptation), they tend to synchronize across large brain regions. This large-range synchronization is an emerging property and occurs despite the fact that the different connections have a range of time delays. It was also shown that in this synchronized slow-wave state, the functional connectivity is closer to the structural connectivity, compared to the asynchronous mode (Goldman et al., 2023, 2020; Sacha et al., 2025).

A review of the neural correlates of consciousness examined various brain oscillations, such as the well-known gamma oscillations, and concluded that the best correlate of consciousness was the desynchronized EEG activity (Koch et al., 2016). This corresponds to the asynchronous mode, as simulated in Figure 5. It was also shown that in the awake and conscious brain, the functional connectivity strongly differs from the structural connectivity, while in the sleeping or anesthetized brain, when displaying slow waves, the functional connectivity is much closer to the structural connectivity (Barttfeld et al., 2015). These features were reproduced by the TVB-AdEx whole brain model (Goldman et al., 2021). This shows that the awake brain produces new patterns of activity, as also found at the cellular assemblies level (Filipchuk et al., 2022), while the anesthetized brain tends to produce patterns that more passively follow the structural connections.

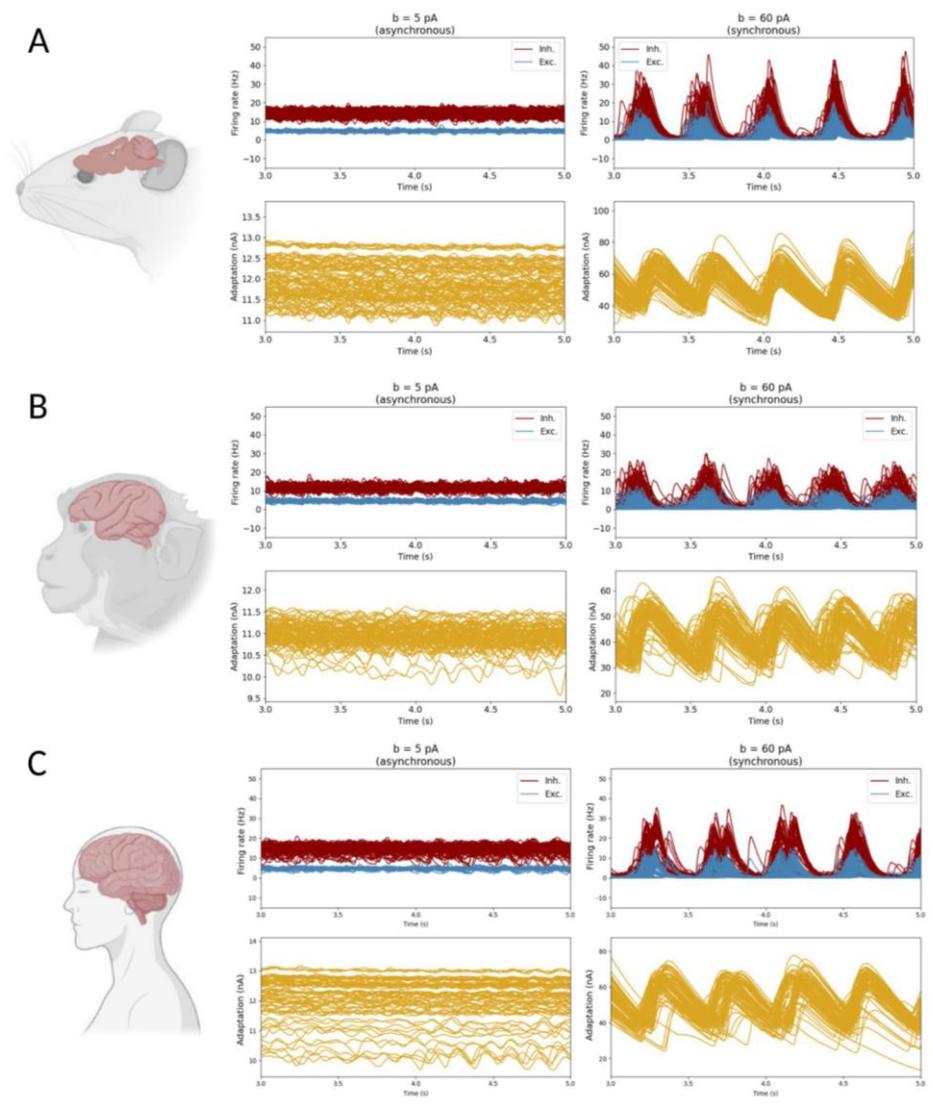

*Figure 5. **Whole-brain simulations of synchronous and asynchronous states.*** *Whole-brain simulations of synchronous and asynchronous states. **(A)** Mouse brain models consisting of 98 nodes each described by the AdEx mean-field model and connected according to the mouse connectome information from the Allen Mouse*

*Brain Atlas. The activity of individual nodes is superimposed, for asynchronous states (left) and synchronized slow-wave oscillations (right).* **(B)** *Similar paradigm simulated with the macaque brain (82 nodes), where the connectivity information was obtained from the Cocomac database.* **(C)** *Asynchronous and synchronized slow-wave states obtained for the human brain (68 nodes). All simulations were done in TVB and are available in EBRAINS. (Adapted from (Sacha et al., 2024)).*

# Spontaneous versus evoked states in humans and clinical implications

## Can slow oscillations occur in the awake brain?

In the last decades, a series of experimental findings using both invasive and non-invasive recordings in humans as well as in animal models have shown that slow oscillations and cortical bistability can also be observed in the awake brain, in particular following sleep deprivation as well as in case of focal brain injuries.

This has been first reported in freely behaving rats (Vyazovskiy et al., 2011) by observing that after a lengthy period of wakefulness, cortical neurons can occasionally show brief periods of silence or Off-periods, just like during sleep, accompanied by slow waves in the local EEG. While some cortical areas may be inactive, other areas remain active, and these periods of "local sleep" increase with the length of the rats' wakefulness. Despite the EEG and behavior still indicating they are awake, the rats' performance on a sugar pellet reaching task were found to decline due to the local populations of neurons falling asleep.

In humans, similar results have been reported by exploiting intracranial LFP and multi-unit recordings collected during presurgical evaluation from *in pharmaco*-resistant epileptic patients while doing a face/nonface categorization psychomotor vigilance task after full-night sleep deprivation (Nir et al., 2017). The authors showed that, immediately before cognitive lapses, selective spiking responses of individual neurons in the medial temporal lobe were weakened, delayed, and prolonged. Additionally, during cognitive lapses, LFPs revealed local slow oscillations that were correlated with impaired single-neuron responses and with baseline theta activity.

Similar results have been obtained in humans non-invasively with high-density EEG, by using individual slow waves in the delta or theta range as a proxy for neuronal slow oscillations and cortical bistability (Huber et al., 2004). These studies have shown that, after sleep deprivation, slow waves increase with time spent awake (time-dependent) selectively in brain regions recruited for long periods of time via a specific task set (use-dependent). Importantly, here again an association has been found between slow waves and behavioral errors. Along the same lines, a recent study has shown that slow waves and behavioral deficits can also occur during wakefulness in the absence of sleep deprivation (Andrillon et al., 2021). By involving a sustained attention task, the occurrence of slow waves was linked to attentional lapses, with slow waves over frontal and posterior areas associated respectively with fast reaction times and more false alarms, and with slower reaction times and misses. The occurrence of slow waves could also predict participants' performance and their subjective experience—including mind wandering (thinking about something other than the task) and mind blanking (thinking about nothing).

Finally, recent studies have shown that a pathological form of sleep-like bistability can intrude in the awake brain of subjects affected by approximately focal brain lesions. The presence of delta waves in the EEG after brain injury has been well known since the beginning of the century (Berger, 1929), but this observation has only very recently been associated with the above-mentioned notion of "local-sleep" by two independent studies showing the presence of fully-fledged sleep-like responses to TMS over the perilesional areas of awake stroke patients (Sarasso et al., 2020; Tscherpel et al., 2020; for a review see Massimini et al., 2024). Along the same lines, pathological sleep-like slow waves have been recorded with invasive recordings in epileptic patients after controlled surgical lesions by radiofrequency-thermocoagulation (Cossu et al., 2015) performed for epilepsy treatment (Russo et al., 2021). This study was able to overcome challenges typical of research into stroke, such as low resolution of non-invasive recording and a lack of baseline recordings pre-injury. This unique approach based on radiofrequency-thermocoagulation allowed for the electrophysiological characterization of perilesional cortical bistability and its effects on distant nodes. Slow oscillations were prominent within a 28 mm radius of the lesion and extended to 60 mm away. This percolation did not follow a distance rule but instead was predicted by individual patterns of long-range effective connectivity. This provides evidence for the hypothesis of remote functional disruption in regions connected to the lesion site.

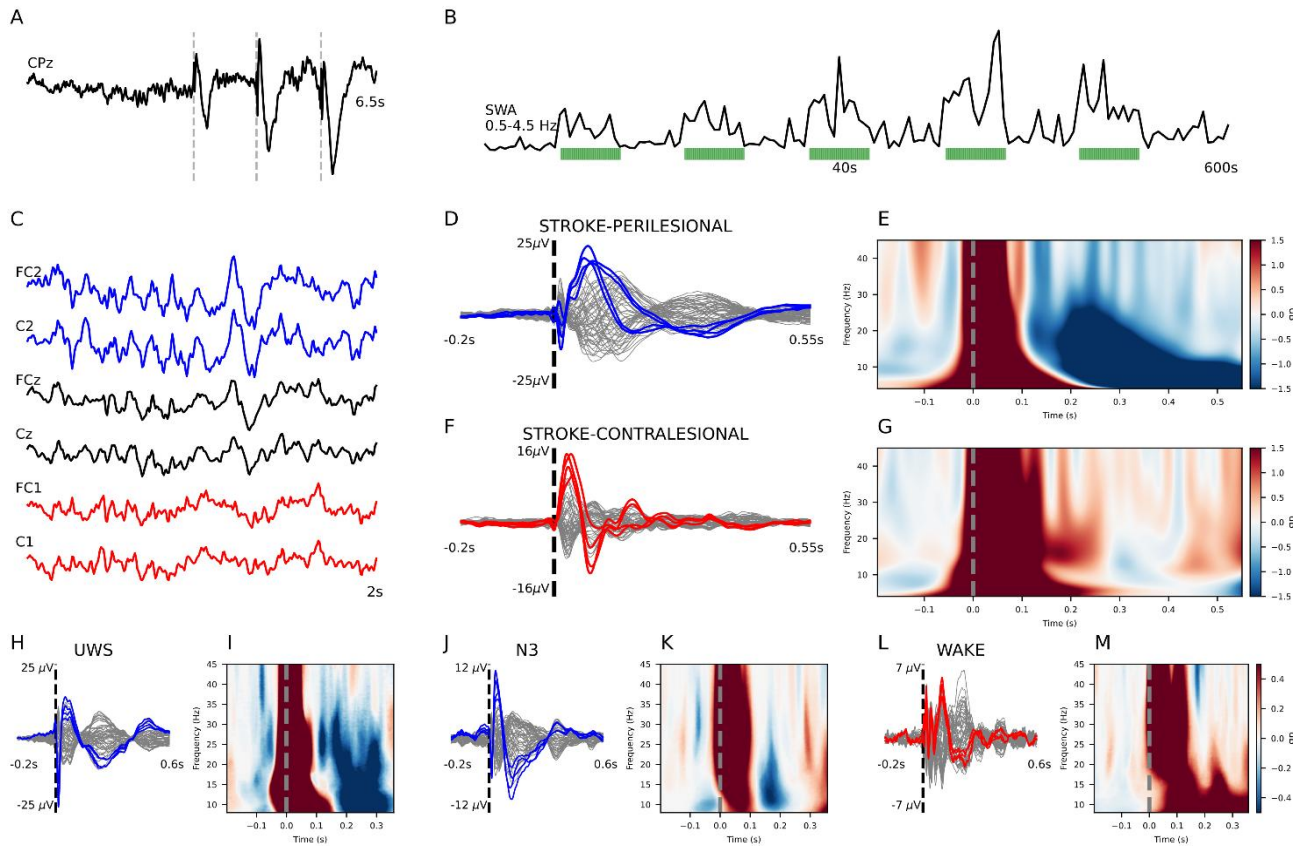

*Figure 6. From spontaneous activity to perturbation. (A)* Time course of one channel of scalp EEG during N3 sleep showing spontaneous activity and single trial responses to TMS (vertical dashed lines). *(B)* Time course of average Delta activity across all EEG channels during periods of spontaneous activity and TMS (green horizontal bars). *(C)* Spontaneous activity from four EEG channels in the perilesional area of a stroke patient. *(D)* Responses to TMS when stimulating the perilesional area of the same patient shown in panel c. Highlighted channels correspond to the four channels showing the largest responses and correspond to those shown in panel c. *(E)* Time-frequency representation of the responses to TMS of the four channels highlighted in panel D. F, G, H. Same as panels C, D and E, respectively, but from a contralesional area. I, J, K, L, M. Responses to TMS and time-frequency representations of the four highlighted channels for a UWS patient, and a neurotypical subject during N3 sleep and wakefulness.
34...

# Perturbation can reveal cortical bistability regardless of ongoing activity

The observational approaches described previously can provide profound, but ultimately correlative, insights into the description of cortical bistability and the spontaneous alternation between On- and Off-periods (i.e. Up and Down states). To parallel observational approaches, neuronal/cortical bistability can be studied by a causal perspective by employing a perturbational approach (Massimini et al., 2009). The role of the perturbation is to transiently alter spontaneous activity (often engaging activity-dependent mechanisms) to trigger an Off-period, thus unveiling the presence of underlying bistability ((Massimini et al., 2007); Figure 6A)). Indeed, cortical perturbations, either electrical or magnetic, are often capable of evoking clear-cut Off-periods, which are not otherwise present in spontaneous activity not only during N2 sleep (Pigorini et al., 2015) but also in stroke and traumatic brain injury patients (Rosanova et al., 2018; Sarasso et al., 2020) (Figure 6). Similarly, perturbations can reveal bistable dynamics during anesthesia both in human (Sarasso et al., 2014b) and in animal models *in vivo* (Arena et al., 2021; Dasilva et al., 2021) and *in vitro* (D'Andola et al., 2018). The data collected during this set of experiments are reviewed in detail in the companion manuscript associated with the present study (Destexhe et al., 2025). Overall, these experimental findings (reported in Fig. 5A and in the dataset) can be mechanistically explained by means of simulations and mean-field theory of cortical modules endowed with activity-dependent adaptation (Cattani et al., 2023; Mattia and Sanchez-Vives, 2012). Indeed, this model points to a key role of activity-dependent adaptation mechanisms in shaping responses to perturbation and in affecting the complexity of cortico-cortical interactions. These modeling results provide a general theoretical framework and a mechanistic interpretation for experimental data that are relevant for both physiological states and brain injured patients.

# Conclusions

In this review, we have argued that cortical brain states—in particular the extremes of synchrony and asynchrony—can only be fully understood through a multiscale perspective that integrates cellular, circuit, macroscopic, and whole-brain dynamics. We have shown that cortical networks generate slow (<1–4 Hz) alternations of Up and Down states across scales—from isolated slices to the intact human brain under non-REM sleep or deep anesthesia—revealing a fundamental bistable attractor state and a default activity pattern of cortical circuits.

Slow waves propagate as coherent travelling patterns whose repertoire and entropy expand as wakefulness is approached. Electrophysiological recordings and computational models demonstrate that the spatial diversity of wavefronts reflects underlying changes in local excitability, and that these changes can be captured by variations in synaptic drive and spike-frequency adaptation, without rewiring of structural connectivity

Awakening, emergence from anesthesia, and recovery of consciousness all coincide with a transition from slow wave synchrony toward sustained higher frequency, low amplitude, asynchronous firing. This shift reverses the spectral slope of the EEG, lowers cortical bistability, unlocks richer functional connectivity patterns and enhances complexity.

Causal probing (TMS-EEG or intracranial stimulation) further confirms that synchronous states support simple, local evoked responses, whereas the awake cortex generates complex, deterministic cascades of activity, suggesting that cortical bistability is both a marker and a mechanism of unconsciousness across sleep, anesthesia and brain injury.

By unifying experimental and computational approaches—we provide a framework to predict how neuromodulatory or pharmacological manipulations will shift cortical dynamics as well as to interpret electrophysiological biomarkers of consciousness in clinical settings in close connection with the underlying cellular mechanisms.

We conclude that brain states are shaped by local circuit mechanisms, large-scale connectivity and neuromodulation. A unified multiscale theory of cortical dynamics is relevant not only for understanding consciousness levels but also for improving patient care through monitoring and control of brain states.

# Acknowledgements

This project/research has received funding from the European Union's Horizon 2020 Framework Programme for Research and Innovation under the Specific Grant Agreement No. 945539 (Human Brain Project SGA3).

**The EBRAINS Collaboratory where Live Figures are hosted is:**
https://wiki.ebrains.eu/bin/view/Collabs/live-figures-multiscale-brain-states/ **in the Drive tab.**

- The links to the notebooks for reproducing the Figures are available on the entry page of the EBRAINS Collaboratory. These links take the reader directly to a JupyterLab instance where the corresponding code and data are accessible.

- An EBRAINS account is needed to run the notebooks at lab.ebrains.eu.

Some of the datasets have been curated and made publicly available in the EBRAINS Knowledge Graph (https://search.kg.ebrains.eu/)

| Dataset | URL |
|---|---|
| Transitions between different naturally and artificially-induced brain states in in vivo rats | https://search.kg.ebrains.eu/instances/e07ab90a-6308-469c-8bc7-a41234103ad3 |
| Propagation modes of slow waves in mouse cortex | https://search.kg.ebrains.eu/instances/7866daf2-7064-4fa0-b6a2-0b1c899ba35f |
| Spontaneous cortical activity in human and ferret brain slices (in vitro) | https://search.kg.ebrains.eu/instances/802fcf0c-ba03-47f9-9051-6143600b980a |
| Slow waves form expanding, memory-rich mesostates steered by local excitability in fading anesthesia | https://search.kg.ebrains.eu/instances/39526a60-0c72-4b20-b850-a2bffbf02049 |
| Interactive Exploration of Brain States and Spatio-Temporal Activity Patterns in Data-Constrained Simulations | https://search.kg.ebrains.eu/instances/3ebdd555-f965-477c-8a0e-4c220014d138 |
| Coregistration of simultaneous HD-EEG and intracranial EEG during single pulse intracerebral stimulation in wakefulness and sleep | https://search.kg.ebrains.eu/instances/a3e9cd95-d601-40ed-b5fa-e5a9fd01005a |